\documentclass[namedreferences]{solarphysics}
\usepackage[hyperref,optionalrh,solaromanenum]{spr-sola-addons} 
\usepackage{graphicx}                    
\usepackage{graphics}  
\usepackage{amssymb}                    
\usepackage{color}                       
\usepackage{breakurl}                         

\newcommand{\arcsec}{\mbox{\ensuremath{\,\,\!\!^{\prime\prime}}}}
\newcommand{\farcs}{\mbox{\ensuremath{.\!\!^{\prime\prime}}}}

\begin{document}

\begin{article}

\begin{opening}

\title{Kinematics and Magnetic Properties of a Light Bridge in a Decaying Sunspot}

%

\author[addressref={aff1},corref,email={mfalco@oact.inaf.it}]{\inits{M. Falco}\fnm{M. Falco}~\lnm{}}
\author[addressref={aff2},corref,email={}]{\inits{J. M. B.}\fnm{J. M. Borrero}~\lnm{}}
\author[addressref={aff1},corref,email={}]{\inits{S. L. G.}\fnm{S. L. Guglielmino}~\lnm{}}
\author[addressref={aff3},corref,email={}]{\inits{P. R.}\fnm{P. Romano}~\lnm{}}

\author[addressref={aff1},corref,email={}]{\inits{F. Z}\fnm{F. Zuccarello}~\lnm{}}
\author[addressref={aff4},corref,email={}]{\inits{S. C.}\fnm{S. Criscuoli}~\lnm{}}
\author[addressref={aff5},corref,email={}]{\inits{A. C.}\fnm{ A. Cristaldi}~\lnm{}}
\author[addressref={aff5},corref,email={}]{\inits{I. E.}\fnm{I. Ermolli}~\lnm{}}
\author[addressref={aff6},corref,email={}]{\inits{A. S.}\fnm{S. Jafarzadeh}~\lnm{}}
\author[addressref={aff6},corref,email={}]{\inits{L. R.}\fnm{L. Rouppe van der Voort}~\lnm{}}

%
\runningauthor{Falco et al.}
\runningtitle{Kinematics and Magnetic Properties of a Light Bridge in a Decaying Sunspot}

\address[id={aff1}]{Dipartimento di Fisica e Astronomia - Sezione Astrofisica, Universit\`{a} di Catania, via S. Sofia 78, 95123 Catania, Italy}
\address[id={aff2}]{Kiepenheuer-Institut f\"{u}r Sonnenphysik, Sch\"{o}neckstr (KIS). 6, D-79110, Freiburg, Germany}
\address[id={aff3}]{Istituto Nazionale di Astrofisica (INAF) -- Osservatorio Astrofisico di Catania, via S. Sofia 78, 95123 Catania, Italy}
\address[id={aff4}]{National Solar Observatory (NSO), Sacramento Peak Box 62, Sunspot, NM 88349, USA}
\address[id={aff5}]{Istituto Nazionale di Astrofisica (INAF) -- Osservatorio Astronomico di Roma, Via Frascati 33, I-00078 Monte Porzio Catone, Italy}
\address[id={aff6}]{Institute of Theoretical Astrophysics, University of Oslo, P.O. Box 1029 Blindern, N-0315 Oslo, Norway}
\begin{abstract}

We present the results obtained by analyzing high spatial and spectral resolution data of the solar photosphere acquired by the CRisp Imaging SpectroPolarimeter at the Swedish Solar Telescope on 6 August 2011, relevant to a large sunspot with a light bridge (LB) observed in NOAA AR 11263. These data are complemented by simultaneous Hinode Spectropolarimeter (SP) observation in the Fe I 630.15 nm and 630.25 nm lines.
The continuum intensity map shows a discontinuity of the radial distribution of the penumbral filaments in correspondence with the LB, which shows a dark lane (about 0.3'' wide and about 8.0'' long) along its main axis. The available data were inverted with the Stokes Inversion based on Response functions (SIR) code and physical parameters maps were obtained. The line-of-sight (LOS) velocity of the plasma along the LB derived from the Doppler effect shows motions towards and away from the observer up to 0.6 km/s, which are lower in value than the LOS velocities observed in the neighbouring penumbral filaments. The noteworthy result is that we find motions toward the observer up to 0.6 km/s in the dark lane where the LB is located between two umbral cores, while the LOS velocity motion toward the observer is strongly reduced where the LB is located between an umbral core at one side and penumbral filaments on the other side. Statistically, the LOS velocities correspond to upflows/downflows and
comparing these results with Hinode/SP data, we conclude that the surrounding magnetic field configuration (whether more or less inclined) could have a role in maintaining the conditions for the process of plasma piling up along the dark lane. The results obtained from our study support and confirm outcomes of recent magnetohydro-dynamic simulations showing upflows along the main axis of a LBs. 
\end{abstract}

%

\end{opening}

%
\section{Introduction}

In recent years, our understanding of the physical mechanisms responsible for the formation and evolution of sunspots has been greatly improved thanks to high temporal-, spatial- and spectral-resolution polarimetric data. These observations have unveiled some physical properties of several fine structures of the sunspots \citep{ThomWeiss, Borrero_2011, Rempel_2011} such as umbral dots (inside the umbra), dark-core filaments (in the penumbra) and light bridges (LBs, separating some umbral portions).

The study of the LBs plays an important role for understanding the grower and decay phases of sunspots. LBs are bright and elongated structures delineating the border between dark umbral cores. In some cases they form during the coalescence of sunspots, while in other cases they are evidence of re-establishment of granular motions within the spot, and often indicate the beginning of spot fragmentation \citep{Vazquez73}. Moreover, before the formation of a LB, several umbral dots emerge in the location where a LB will be formed, and the LB structure rapidly intrudes from the leading edge of penumbral filaments into the umbra \citep{Katsukawa_2007}.

According to \cite{ThomWeiss}, there are two types of LBs: ones segmented along their length, with bright segments resembling tiny granules separated by narrow dark lanes oriented perpendicular to the axis of the LB \citep{berger03}; and others, which are unsegmented, and, according to \citet{Lites04}, resemble the elongated features seen in the penumbra. Moreover, LBs are classified as faint light bridges (FLBs), which are narrow bright features inside the umbra, and as strong light bridges (SLBs), separating different umbral fragments \citep{Sobotka1994}.
LBs often exhibit a granular morphology, even if the size, lifetime, and brightness of these granules are different from those found in the granulation of the quiet photosphere.
In particular, the granular structures forming SLBs are generally smaller than the ones in the quiet photosphere: the typical sizes being $1.2''$ \citep{Sobotka1994}, instead of $1.5''$. The intensity of the LB granules is about 85\% of the mean photospheric intensity. By following the temporal evolution of these sub-structures during their irregular motions inside the LB, proper motions with velocities up to $1.5\,\mathrm{km\,s}^{-1}$ have been detected \citep{Hirz2002}. The lifetime distribution of these granular features shows a maximum at 5 min and a second peak at approximately 20 min \citep{Hirz2002}.

In LBs the magnetic fields have a lower strength and are sparser and more horizontal than in the neighbouring umbrae. LBs with their weak and inclined fields therefore represent a discontinuity in the regular umbral field \citep{Lites91, Leka97}. Recent observations by \citet{Jurcak05} and \citet{Jurcak06} pointed out an essentially field-free region at the deepest visible level of two LBs, but with magnetic canopies spreading from either sides of the LBs and merging above them.

Many segmented LBs also show a narrow dark lane along their main axis \citep{Rimmele08}. \citet{berger03} showed that this lane has a typical width that varies from $0.2''$ to $0.5''$: in the larger section the features (the granules along the sides of the lane) resemble large-scale modified convection, while in the smaller sections of the LB, the granules appear increasingly smaller, until only a central dark lane is observed, probably due to the alignment of convection cells caused by the magnetic field. Often during its lifetime, portions of the dark lane dissolve and then reform again \citep{Rimmele08}. 

An analysis of a LB velocity field carried out by \citet{Rimmele97} provided evidence of the presence of sinking plasma in the axial channel. From the correlation between vertical velocity and continuum intensity in the granules belonging to a LB, \citet{Rimmele97} could confirm their convective origin. In this regard, it is important to find out whether the motions observed in LBs have a magneto-convective origin or are due to convection penetrating from the sub-photospheric layers into a field-free gap \citep{ThomWeiss}. Furthermore, this study is also important to understand the physical processes at the basis of the sunspot formation. Actually, there are two theoretical models which attempt to describe the sunspot formation process: the monolithic model \citep{Cowling_1957} and the spaghetti-like model \citep{Parker_1979a,Parker_1979b,Parker_1979c}, but according to \citet{Rempel_2011} it seems that the monolithic model is the favoured one.

More recent observations show that in segmented LBs the dark lane is characterized by upflows, while in the intergranular lanes of normal granulation we can see downflows \citep{vander10}. Therefore, the process which originates the dark lane seems to be plasma pile-up caused by the strong decrease of the vertical upflow near the surface. The plasma is forced by the cusp-like surrounding magnetic field into a region with enhanced density and therefore larger opacity \citep{Giordano2008}. This causes the elevation of the $\tau$ = 1 (where $\tau$ denotes the optical depth) surface into a cooler, higher part of the atmosphere resulting in a dark lane in intensity images \citep{Sch06}. Moreover, recent observations have also revealed that some types of LBs are accompanied by remarkable long-lasting plasma ejections or surge activities in the chromosphere \citep{Asai01, Shimizu_2009, Schimizu11, Louis2014, Toriumi_a}.

In this paper we provide a further observational contribution to the understanding of plasma motions and magnetic fields in a LB, by using a very high quality dataset acquired with the \textit{CRisp Imaging SpectroPolarimeter} on the 1-m \textit{Swedish Solar Telescope} on $6$ August 2011, imaging NOAA AR 11263. In the next section we describe the dataset analysed and the methods applied. In Section 3 we report our results, in Section 4 we discuss their interpretation, and in Section 5 we draw our conclusions.

\section{Observations and Data Analysis}

The active region NOAA AR 11263 was observed on $6$ August 2011 at N16W43 ($\mu=0.76$, which is the value of the cosine of the heliocentric angle $\theta$ of the observations) using the \textit{CRisp Imaging Spectropolarimeter} \cite[CRISP;][]{Scharmer08} mounted at the 1-m \textit{Swedish Solar Telescope} \cite[SST;][]{Scharmer_2003}, during a joint observing campaign (HOP 0195) with the \textit{Hinode} satellite \citep{Kosugi_2007}. The CRISP spectropolarimetric measurements were taken from 09:53:32 UT to 10:48:43 UT along the Fe I line pair at 630.15 nm  and 630.25 nm with 15 spectral points for each line, in steps of 4.4 pm from $-26.8$ pm to $34.8$ pm for 630.15 nm and steps of $4.4$ pm from $-30.8$ pm to $30.8$ pm for 630.25 nm with respect to the line center of each line. \citet{Cristaldi2014} used the same kind of dataset for a different target observed during this observing campaign. The average cadence of each scan was 28 seconds. Liquid crystals modulated the light cycling through four polarization states (\textit{I}, \textit{Q}, \textit{U}, and \textit{V}) \citep{Schnerr_2011}, and 5 exposures per polarization state were acquired, resulting in a total of 20 exposures per line position. The pixel size of the CRISP cameras was $0.06''$ pixel$^{-1}$ at 630.15 nm and 630.25 nm. The noise level is different for each Stokes parameter: $4\times10^{-3} I/I_{c}$ (where $I_{c}$ is the mean quiet Sun continuum intensity value) for Stokes \textit{Q}, $3\times10^{-3} I/I_{c}$ for Stokes \textit{U}, and $8\times10^{-3} I/I_{c}$ for Stokes \textit{V}. The field of view (FOV) of these SST measurements was $57.5'' \times 57.3''$.

Moreover, the dataset acquired by Hinode satellite during the joint observing campaign was used to obtain information on the magnetic field topology of the AR. The \textit{Hinode}/SP \citep{Tsuneta_2008, Lites_2013}, recorded the Stokes profiles along the Fe I line pair at 630.15 nm and 630.25 nm with a pixel sampling of $0.317''$ and a noise level of about $10^{-3} I/I_{c} $ (fast mode). Level 2 data obtained from the Milne-Eddington gRid Linear Inversion Network (MERLIN) code \citep{Lites_2007} were used in our analysis. We applied the non-potential magnetic field calculation technique \citep[NPFC;][]{georg05} to the inverted dataset to perform azimuth disambiguation in solar vector magnetograms, obtaining inclination and azimuthal angle maps in the local solar frame.

To follow the global evolution of the AR and the formation of the LB we also analyzed continuum images and line-of-sight (LOS) magnetograms in the Fe I line at 617.3 nm from $2$ to $7$ August 2011 taken by the \textit{Helioseismic and Magnetic Imager} \citep[HMI;][]{Scherrer2012} onboard of the \textit{Solar Dynamics Observatory} \citep[SDO;][]{pesnell12}. Both datasets were characterized by a pixel size of $0.5''$ and a time cadence of 2 hours. 

The SST data have been processed using the Multi-Object Multi-Frame Blind Deconvolution \citep[MOMFBD;][]{vannoort05} technique in order to achieve the highest spatial resolution in combination with the adaptive optics. For the data processing we followed the different steps in the CRISPRED reduction pipeline for CRISP data \citep{jaime15}. Wideband images, acquired simultaneously with the spectro-polarimetric scans, have been used as a so-called anchor channel in the reduction procedure to ensure precise alignment between the sequentially recorded CRISP narrowband images. Following the data calibration, we obtained two three-dimensional datacubes containing restored, aligned data with a high angular resolution of $0.16''$.
The blueshift variation from the center of the FOV towards the edge, due to the Fabry-P\'erot system, was corrected during the reduction procedure. 

We applied the Stokes Inversion based on Response functions \citep[SIR;][]{RuizIniesta92} code to the SST sequence acquired at 10:17 UT, obtained during the best seeing condition, to obtain maps of the magnetic field strength and temperature in a sub-array FOV centered on the LB region. Using the SIR code we inverted simultaneously the spectra acquired in both the lines of the Fe I line pair at 630.1/630.2 nm. Stokes \textit{I} values in the red wing of the Fe I 630.25 nm line were not included in the profile of the inverted pixels because the signal is altered by telluric lines and, for low temperatures, by molecular blends.
We used two different models as initialization of the inversion, depending on the region, identified by a threshold in the continuum intensity, forming the FOV: a penumbral model ($0.4<I/I_{c}<0.8$) and an umbral model ($I/I_{c}<0.4$).
In the penumbral model, we changed the temperature (\textit{T}) and the electron pressure ($p_{e^{-}}$) using the values described by \citet{Del_Toro_1994}, and we used a value of 1000 G for the magnetic field strength, and $1.0\,\mathrm{km\,s}^{-1}$ for the LOS velocity.
For the umbral model we used the \textit{T} and $p_{e^{-}}$ values provided by \citet{Collados_1994} (corresponding to an umbral model for a small spot), and we used a constant value of 2000 G for the magnetic field strength as an initial guess.
The temperature stratification of each component was modified with two nodes.

The other physical parameters (magnetic field, LOS velocity, inclination and azimuth angles) were assumed to be constant with height (number of nodes equal to one). We used thus Milne-Eddington-like approximation to provide estimation of the average physical parameters over the range of line formation heights.
A fixed macroturbulence velocity of $2.95\,\mathrm{km\,s}^{-1}$ was used in order to mimic the effects of the spectral point spread function (PSF) of the instrument. A fixed filling factor of one was used for the inversion.
The straylight contamination was not considered during the inversion owing to the fact that the Stokes \textit{I} profile in the Fe I line at 630.2 nm is heavily affected by blends, thereby severely affecting our ability to determine its contribution. We decided to apply more weight to \textit{Q}, \textit{U}, and \textit{V} Stokes  parameters, by a factor of four with respect to Stokes \textit{I} owing to the larger noise in the intensity than in the polarization profiles. 
Therefore, due to the uncertainties that could affect the velocity values derived from the SIR inversion, we decided to derive the Doppler velocity by applying a Gaussian fit to the Fe I line profile at 630.15 nm. Moreover, the CRISP dataset was not used to study the inclination angle because the LB's FOV was not sufficient to apply the nonpotential magnetic field calculation (NPFC) code. 

To study the kinematics in the LB region, we obtained the Doppler velocity of plasma motions by applying to five sequences of the SST dataset a Gaussian fit to the Fe I line profile at 630.15 nm with the MPFIT routine \citep{Markwardt_2009} in Interactive Data Language (IDL). The values of LOS velocity were deduced from the Doppler shift of the centroid of the fitted line profiles in each spatial point.
The LOS velocity map was calibrated by subtracting the mean velocity of the pixels of the umbra in the FOV (pixels which have a threshold of the continuum intensity lower than 0.4 $I/I_{c}$), assuming that the umbra of the sunspot was at rest, according to \citet{Balthasar}.
We estimated the uncertainty affecting the velocity measurements considering the standard deviation of the centroids of the line profiles estimated in all points of the FOV. Thus, the estimated relative error in the velocity is $\pm0.2\,\mathrm{km\,s}^{-1}$.
Moreover, we remind the reader that, given the position of AR 11263 at N16W43, the measured LOS velocities do not correspond to velocities perpendicular to the solar surface.

\section{Results}

Figure \ref{fig1} shows the HMI/SDO continuum image and the corresponding magnetogram taken on $6$ August 2011 for NOAA AR 11263. On this day the AR is characterized by a preceding main negative polarity sunspot showing a light bridge and by several smaller positive polarity ones (see left and right panels of Figure \ref{fig1}).

Figure \ref{fig2} shows the evolution of the preceding spot from $3$ to $6$ August 2011. The well-defined umbra and penumbra (Figure \ref{fig2}, top left panel), start to fragment on $4$ August, when the penumbral filaments in the north-western part of the spot seem to penetrate in the umbral region (see the red contour in Figure \ref{fig2}, top right panel). On $5$ August (Figure \ref{fig2}, bottom left panel) the preceding sunspot exhibits two umbrae with different shapes, one more elongated and the other with a more circular shape, inside the same penumbra. The LB appears and completes its formation on $6$ August, before the start of the CRISP observations analysed in the following (the CRISP's FOV is indicated by the box in Figure \ref{fig2}, right bottom panel).

Figure \ref{fig3} shows the continuum intensity map observed by CRISP on $6$ August at 10:17 UT, with the LB oriented approximately along the north-south direction. It is worth noting that the radial distribution of the penumbral filaments is modified in the regions near the LB. In fact, penumbral filaments south of the LB are smaller than the others around the sunspot. Moreover, in this area the photospheric granulation seems to prevail over the penumbra. Further, to the north-east of the LB we note some small dark regions characterized by a local higher value of the magnetic field strength (compare with Figure \ref{fig4}, top panel).

In Figure \ref{fig4} we show the magnetic field strength and inclination angle maps in the local solar frame obtained by the MERLIN inversion on Hinode/SP data. In the bottom panel $0^{\circ}$ and $180^{\circ}$ correspond to radially outward and inward magnetic field, respectively. The magnetic field strength map indicates that in the LB region the magnetic field strength is lower than in the surroundings. The inclination map shows in the region of the LB an inclination angle lower than the magnetic field inclination of the two umbral zones, where it is $\approx 180^{\circ}$. These lower inclination angles are more evident between 210\arcsec\,and 220\arcsec\,in the \textit{y} direction, corresponding to the northern part of the LB.


We restricted our analysis of the LB properties to the data acquired by CRISP/SST at 10:17 UT in the FOV indicated by the box reported in Figures \ref{fig3} and \ref{fig4}.
We divided the LB into two parts, characterized by a different configuration of the magnetic field at its sides: the northern part (corresponding to the upper part of the LB and indicated by LB$_n$ in Figure \ref{fig5} and Figure \ref{fig6}) with the larger umbral core of the spot at the eastern side and the penumbral filaments at the opposite side (compare with Figure \ref{fig3}), and the southern part (corresponding to the bottom part of the LB and indicated by LB$_s$ in Figures \ref{fig5} and \ref{fig6}) with the two umbral cores at both sides.

In Figure \ref{fig5} (top panel) we show a $\approx 90^{\circ}$ rotated zoomed image of the LB region displayed in Figure \ref{fig3} (white box), where the black line indicates the dark lane of the LB.
The narrow dark lane along the main axis of the LB has an average width of $0.3''$ and a length of about $8.0''$. It seems to connect two dark penumbral filaments located at both sides of the LB. Moreover, the LB appears segmented along its length by tiny granules (sizes from $0.2''$ to $0.8''$) separated by narrow dark lanes oriented perpendicularly to the LB axis.
In the other panels of Figure \ref{fig5} we plot the LOS velocity along the dark lane measured around the time of the best seeing sequence (10:17:05 UT). These plots show that, in the LB$_{n}$ region, the LOS velocity values are between $0$ and $-0.2\,\mathrm{km\,s}^{-1}$ (negative velocity values indicate motions toward the observer), while, in the LB$_{s}$ region, the LOS velocity is higher and can reach up to $ -0.8\,\mathrm{km\,s}^{-1}$.
Therefore, we note that the dark lane shows mostly motions toward the observer, which appear to be higher in LB$_s$ where the LB is located between two umbral regions.
We can see that this trend of the LOS velocity along different portions of the LB persists at least for 4-5 minutes, i. e., for a lifetime comparable with the lifetime of the LB granules.

Therefore, in order to investigate how the configuration of the magnetic field at the LB sides can influence the dark lane properties, we studied the intensity, the magnetic field strength, the temperature, and the LOS velocity inside and around the LB at 10:17:05 UT (see Figure \ref{fig6}) with particular attention to the vertical segments reported in Figure \ref{fig5} (bottom panel), where the LOS velocities of the dark lane are completely different.
 
The granules at the western side of the LB$_n$ (see the white arrow in the LB$_n$ portion in Figure \ref{fig6}, top left panel) are larger than the ones at the eastern side (conversely in the southern part of the LB). The maps of the magnetic field strength and of the temperature at $\log(\tau_5)=0.0$ (where $\tau_5$ is the optical depth at 500.0 nm) obtained by the SIR inversion (Figure \ref{fig6}, top right and bottom left panels) indicate that the granules of the LB that are characterized by a larger size correspond to regions with a weaker magnetic field and a higher temperature. The LOS velocity of the plasma along the LB is on average lower in value (of the order of $\approx \pm0.6\,\mathrm{km\, s}^{-1}$) than the LOS velocity in the neighbouring penumbral filaments, where the LOS component is of the order of $\approx 2\,\mathrm{km\,s}^{-1}$ due to the Evershed flow (note that in Figure \ref{fig6}, bottom right panel, 
the LOS velocities are saturated between $-1$ and $+1\,\mathrm{km\,s}^{-1}$ to make more visible the LOS velocity values along the LB).

In Figure \ref{fig7} are analyzed the intensity, the magnetic field strength, the temperature, and the LOS velocity along the two segments perpendicular to the LB drawn in Figures \ref{fig5} (bottom panel) and \ref{fig6}. The plots in the left and in the right columns of Figure \ref{fig7} correspond to the above physical quantities along the segments in the northern and in the southern part of the LB, indicated in Figure \ref{fig6} by labels 1 and 2, respectively. The vertical line in the plots indicates the location of the dark lane.

We note that the intensity of the dark lane is about $0.8\,I/I_{c}$ for both parts of the LB (see Figure \ref{fig7}, top panels). In the LB$_n$ (Figure \ref{fig7}, top left panel) we see that at both sides of the dark lane there are two maxima in the continuum intensity. They correspond to the granules of the LB$_n$ along the line 1. The largest granule has a size of about $0.8''$ and is characterized by a magnetic field strength between 700 G and 1000 G (see Figure \ref{fig7}, second left panel). The smallest granule has a size of about $0.2''$ and a magnetic field of about 1500 G. The same behavior is detected in the region selected in LB$_s$ (see Figure \ref{fig7}, first two right panels). The temperature at $\log(\tau_5)=0$ in the dark lane is about 6000 K (see Figure \ref{fig7}, third panel from the top), while the larger granules of the LB$_n$ reach a temperature of about 6400 K.

In Figure \ref{fig7} we note that the dark lane in the LB$_{n}$ is almost at rest (see bottom left panel), while the granules at its sides are characterized by motion outward the observer (higher in the larger granule). The dark lane in the LB$_{s}$ (see bottom right panel) shows motion toward the observer of the order of $\approx -0.6\,\mathrm{km\,s}^{-1}$ and motion outward the observer between $0.1$ and $0.4\,\mathrm{km\,s}^{-1}$ in the surrounding region.
Moreover, in Figure \ref{fig7} (bottom right panel) the motions toward the observer exceeding $-0.5\,\mathrm{km\,s}^{-1}$ present a slight shift with respect to the location of the dark lane. We think that this slight shift may be due to a foreshortening effect, taking into account that the AR is far from the central meridian.

In order to study how the inclination angle changes inside and in the surroundings of the LB, we used the data acquired by Hinode/SP at 10:05 UT. In Figure \ref{fig8} the Hinode maps of the continuum intensity and inclination angle, obtained using the MERLIN code, are shown for the region containing the LB. We studied the profile of the magnetic field inclination along the two segments drawn in Figure \ref{fig6} and reported in blue and red colors in Figure \ref{fig8}. The inclination angle along the blue line of Figure \ref{fig9} (which refers to the LB$_{n}$ region, located between an umbral region and a penumbral region) shows that in the eastern umbral region the magnetic field inclination has values around $170^{\circ}$, it reaches values lower than $155^{\circ}$ in the center of the LB$_{n}$ and slightly more vertical values in the western penumbral region. In the LB$_{s}$ (red line in Figure \ref{fig9}), located between two umbral regions, the inclination angle decreases to a minimum in the LB$_{s}$ ($\approx 155^{\circ}$), but at both sides in the umbra the inclination angle is $\geq165^{\circ}$. Given that MERLIN inversions do not provide the errors of the parameters, we estimate the errors of these measurements from the inversion of the Hinode/SP data using another Milne-Eddington-based code \citep[Very Fast Inversion of the Stokes Vector (VFISV);][]{Borrero_VFISV}. The typical range of the standard deviation (1$\sigma$) for the inclination maps is between $2^{\circ}$ and $5^{\circ}$. Thanks to this accuracy, we can also appreciate the small difference in the inclination in the two different regions of the LB.

\section{Discussion}

In order to interpret the LOS velocities found in the LB, that have been shown and analysed in Figures \ref{fig5} and \ref{fig7}, we have to discuss if such motions toward/outward the observer may be related to upflows/downflows. We remember that, in any position different from the disk center ($\cos(\theta) =0$), the LOS velocity is given by

   \begin{equation} 
		v_\mathrm{LOS} = v_z \times \cos(\theta) \pm v_h \times \sin(\theta)
   \end{equation} 

\noindent	
where $v_z$ is the upflow/downflow (vertical) component and $v_h$ is the horizontal component of the velocity, and $\theta$ is the heliocentric angle. To understand which velocity component prevails in the region of the LB, we produced a scatter plot of $I/I_{c}$ and LOS velocity in a region of quiet Sun (contained in the FOV) and in the LB region. Figure \ref{fig10} (top panel) shows that there is a rather clear correlation ($r = -0.467$) in the quiet Sun: brighter points, which correspond to areas occupied by granules that harbor upflows, exhibit motions toward the observer, while darker points, corresponding to dark lanes that harbor downflows, exhibit motions away from the observer. So, in a statistical sense, $v_h$ is null and we can safely assume that LOS velocity in the quiet Sun is indicating upward/downward motions. Then, we checked if a comparable correlation could be found also in the region of the LB. However, in the scatter plot relevant to the LB region shown in Figure \ref{fig10} (bottom panel), where we included the data of the five sequences studied in Figure \ref{fig5} to increase statistics, we note only a very slight negative correlation ($r = -0.106$).
	
The latter result would seem to make us less confident that we can associate LOS velocities to upflows/downflows in the LB region. Nonetheless, we can further observe that, even if we had $v_h \neq 0$, such a residual component can be neglected in our analysis. 
	
As a matter of fact, it is known that LBs and penumbral filaments share a common origin \citep{SpruitScharmer}, thus one would expect that if there are large horizontal velocities in the LB, they should be Evershed-like velocities. However, in the LOS map we do not observe such Evershed-like motions (see Figure \ref{fig6}, bottom-right panel) toward the disk or the limb. Therefore, we are sure that Evershed-like horizontal motions are not affecting our results. Furthermore, we note from Figure \ref{fig3} that the direction of the main axis of the LB is almost perpendicular to the direction of the disk center. Horizontal velocities perpendicular to the disk center do not contribute to the measured LOS velocities, because the angle between the LOS direction and $v_h$ is $90^{\circ}$ in such a configuration. Any residual $v_h$ should be oriented along a direction parallel to the main axis of the LB, giving no contribution to the measured LOS velocity.

We can conclude that the LOS velocities that we studied have a vertical velocity component which prevails over the horizontal component and, then, we can refer to those velocities as upflow and downflow plasma motions also in the region of the LB.

\section{Summary and Conclusions}

Understanding the interplay between plasma convection and the magnetic field distribution in sunspot fine structure is fundamental in order to have useful hints on the magneto-hydrodynamical processes occurring in the solar atmosphere and in the underlying layers, hidden from direct observations.

In this framework, this study is aimed at providing a contribution to the understanding of the magnetic and kinematic properties of small-scale features observed in LBs. To accomplish this goal, we analyzed CRISP data for a sequence acquired on $6$ August 2011 at 10:17 UT along the Fe I line profile (630.15 nm and 630.25 nm).
The FOV of interest contains the preceding, negative polarity sunspot of AR NOAA 11263 where, according to HMI/SDO data analysis, a LB had formed in the previous days. 
The sunspot was in the decaying phase and the presence of a LB is evidence of the re-establishment of the granulation within the spot. The LB is quite defined and presents a granular morphology, with a dark lane along all the length of its main axis; according to \citet{Sobotka1994}, this LB is a strong light bridge (SLB).

From the investigation of the continuum intensity images of the sequence acquired by CRISP at 10:17 UT, during the best seeing conditions, we confirm that the LB was segmented: it was in fact characterized by a central dark lane having at its sides granules of different sizes separated by tiny intergranular lanes. The LB was an interesting target because it was characterized by a different configuration at its sides along its length: the southern part was located between two umbral cores, while the northern part was located between an umbral core at one side and penumbral filaments on the other side.

Using the CRISP data, we analyzed the thermal and kinematic properties separately for the upper and lower parts of the LB located between the umbral and the penumbral filaments (LB$_n$) and between the dark umbral cores (LB$_s$), respectively.
In the LB$_n$ there are grains of different size along the dark lane: grains on the western part have a size of about $0.8''$ and magnetic field of $\approx$ 700 G, while the grains on the eastern part have a size of about $0.2''$ and magnetic field of $\approx$ 1500 G. In the LB$_s$ the larger ones are the grains on the eastern side of the dark lane.
The width of the dark lane is about $0.3''$, in agreement with the results obtained by \citet{berger03}.
Furthermore, we found that the intensity of the LB granules is about 0.9 $I/I_{c}$.

The results of this analysis, shown in Figure \ref{fig7}, indicate that in both portions of the LB the dark lane has an intensity of 0.8 $I/I_{c}$ and a temperature at  $\log(\tau_5)=0$ of about 6000 K.

\citet{Nord_2010} found a cusp-shaped central dark lane in a LB formed in a three-dimensional MHD simulation of a field-free gap surrounded by an umbral-like atmosphere and predicted that the dark lane in sunspot LBs harbors upflows. The confirmation of these outcomes derived from numerical simulations with observations is challenging. High spatial resolution observations have shown in fact that LBs are highly spatially structured, with convective motions occurring on spatial scales of a few arcseconds. The mixing of information from these features, if not resolved, can thus lead to wrong estimates of the sign of the plasma velocities, thus explaining the discrepancy of results reported in the literature \citep{Rimmele97}.

Moreover, we observed a different behaviour in the plasma motions in the LB$_s$ and LB$_n$, characterized by different magnetic field configurations at the surroundings.
The LB$_s$ is located between two regions with almost vertical magnetic fields (see the red plot in Figure \ref{fig9}). The dark lane in this region hosts motions toward the observer with LOS velocity between $-0.2$ and $-0.6\,\mathrm{km\,s}^{-1}$, as shown also in Figure \ref{fig5}. This last finding is in agreement with the results of \citet{vander10} and confirms the hypothesis that the dark lane is a cusp-like region where the plasma piles up as a consequence of braking of vertical flows. 
On the other hand, the LB$_n$ is located between regions with an almost vertical magnetic field at one side and a more inclined field on the other side (see the blue plot in Figure \ref{fig9}). The dark lane in this region shows weaker motions toward the observer, probably indicating that the process of plasma pile-up in this condition is somehow modified.

Therefore, we conclude that the configuration of the surrounding magnetic field can play an important role not only in the formation of a cusp-like region with enhanced density and corresponding to the dark lane, but also in the vertical upflow usually observed along these structures. The results obtained from our study thus support recent MHD simulations and observations of magneto-convection in sunspot atmospheres.

\begin{acks}
We are grateful to the University of Catania for providing the funds necessary to carry out the Observational Campaign in La Palma. The 1 m \textit{Swedish Solar Telescope} is operated on the island of La Palma by the Institute for Solar Physics (ISP) of Stockholm University in the Spanish Observatorio del Roque de los Muchachos of the Instituto de Astrof\'isica de Canarias.
\emph{Hinode} is a Japanese mission developed and launched by ISAS/JAXA, with NAOJ as domestic partner and NASA and STFC (UK) as international partners. It is operated by these agencies in co-operation with ESA and NSC (Norway). The HMI/SDO data used in this paper are courtesy of NASA/SDO and the HMI science team. Use of NASA's Astrophysical Data System is gratefully acknowledged.
We are grateful to Dr. Rolf Schlichenmaier for the useful discussions and helpful suggestions. 
This research has received funding from the EC 7th Framework Programme FP7/2007-2013 under the grant agreement eHEROES (project n. 284461).
We are grateful to the SOLARNET project for providing the grant for the Young Researches Mobility (project n. 312495).
This work was also supported by the Instituto Nazionale di Astrofisica (PRIN INAF 2014), and by the University of Catania (PRIN MIUR 2012).
\end{acks}

\bibliography{manuscript_biblio}  

\bibliographystyle{spr-mp-sola}

\newpage

\begin{figure}[h]
 \centering
 \includegraphics[trim=0 0 0 0, clip, scale=0.55]{Figure_1a.ps}
 \includegraphics[trim=0 0 0 0, clip, scale=0.55]{Figure_1b.ps}
\caption{Continuum intensity map (left panel) and LOS magnetogram (right panel) of NOAA AR 11263 obtained by the HMI/SDO in the Fe I 617.3 nm line on $6$ August 2011. The red and green contours indicate the umbral and penumbral borders, as derived by applying an intensity threshold set to $I/I_{c}$=0.5 and $I/I_{c}$=0.9, respectively, and where $I_{c}$ is the mean quiet Sun continuum intensity value. North is at the top, west is at the right.}
   \label{fig1}
\end{figure}

\begin{figure}
   \centering
 \includegraphics[trim=0 0 0 0, clip, scale=0.55]{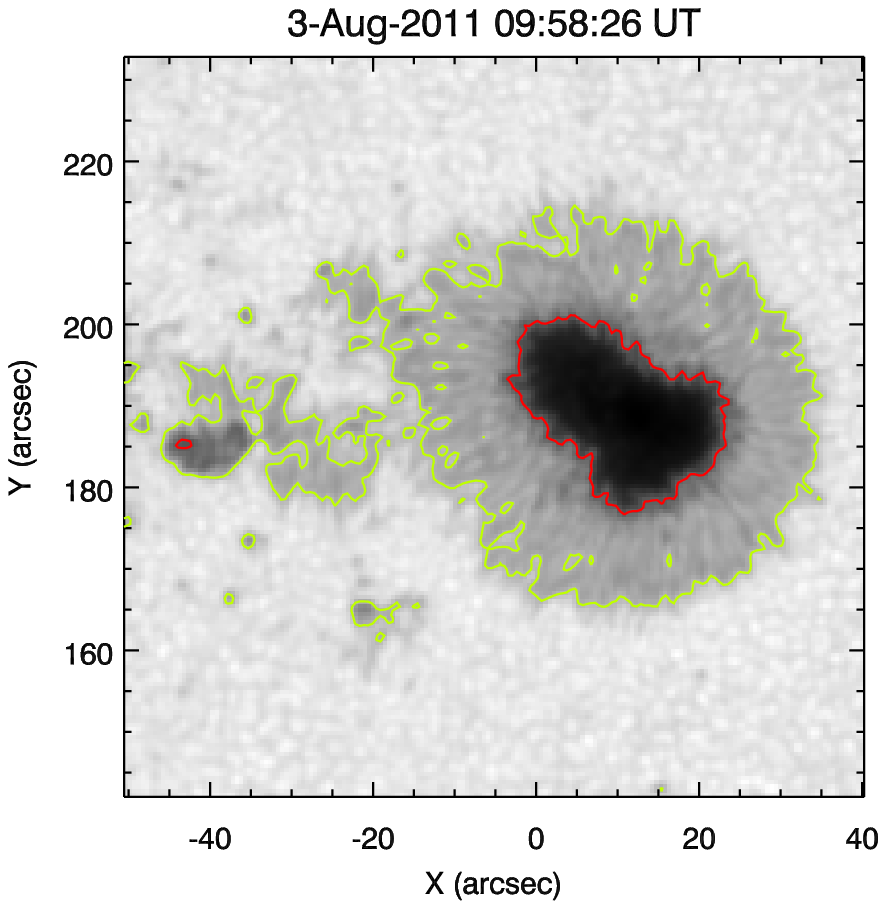}
 \includegraphics[trim=0 0 0 0, clip, scale=0.55]{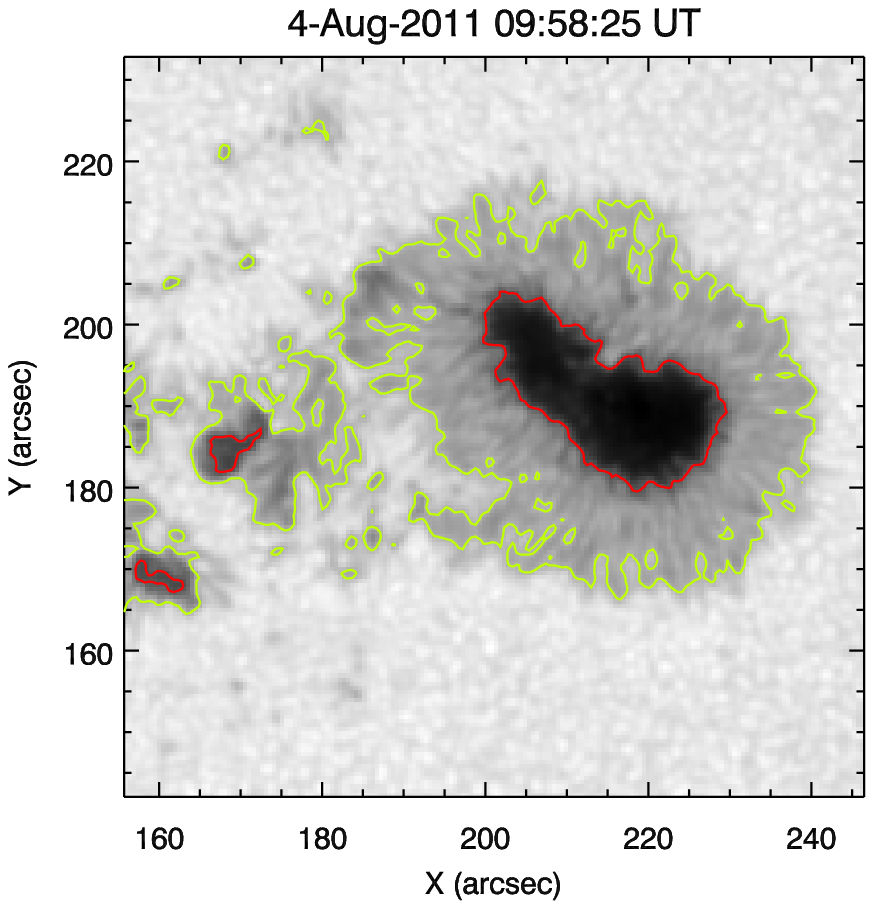}\\
 \includegraphics[trim=0 0 0 0, clip, scale=0.55]{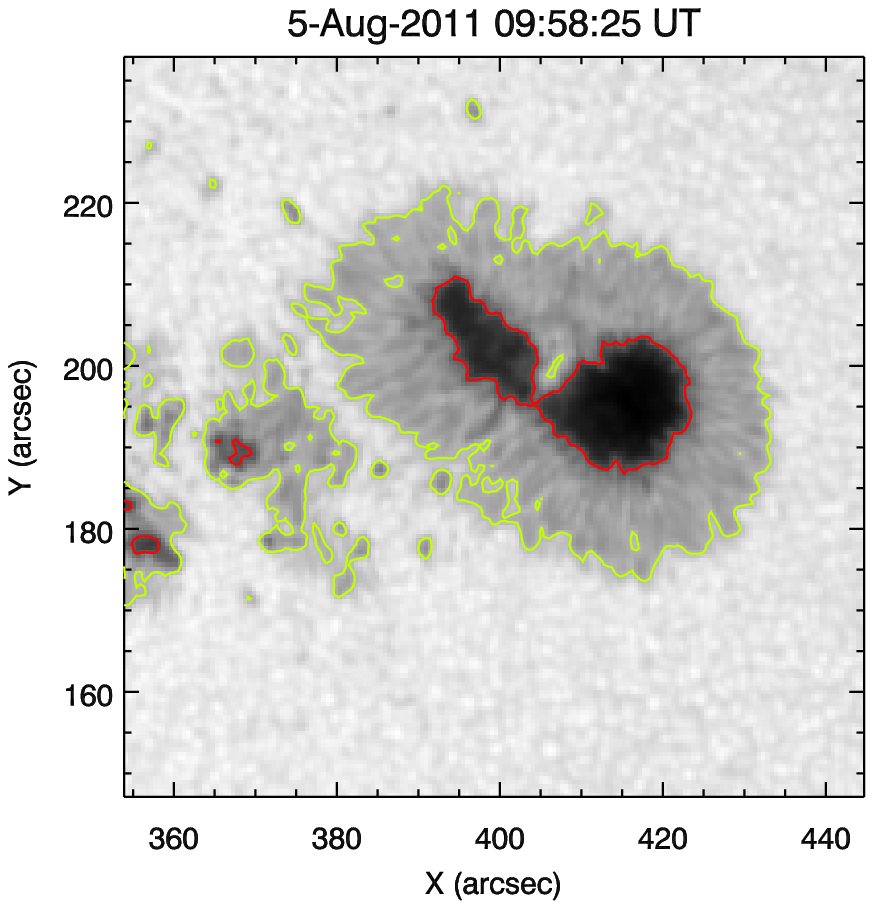}
 \includegraphics[trim=0 0 0 0, clip, scale=0.55]{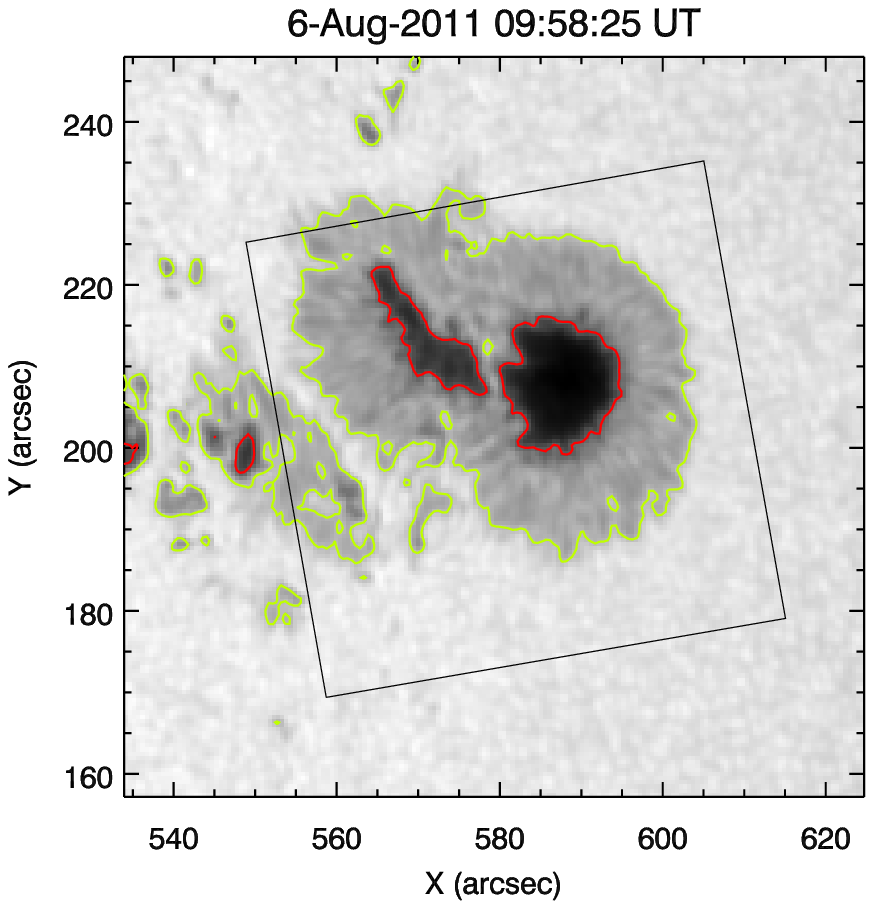}      
 \caption{HMI/SDO images in the continuum of the Fe I line at 617.3 nm of the preceding sunspot of NOAA AR 11263 taken on consecutive days, from $3$ to $6$ August 2011. The FOV is $\approx 90 \times 90$ arcsec. The color contours have the same meaning as in Figure \ref{fig1}. The sequence shows the evolution of the sunspot umbra and the formation of the LB. The box in the bottom-right panel indicates the FOV of the CRISP observations analysed in our study and shown in Figure \ref{fig3}.}
 \label{fig2}
\end{figure}

\begin{figure}
   \centering
 \includegraphics[trim=0 0 0 0, clip, scale=0.6]{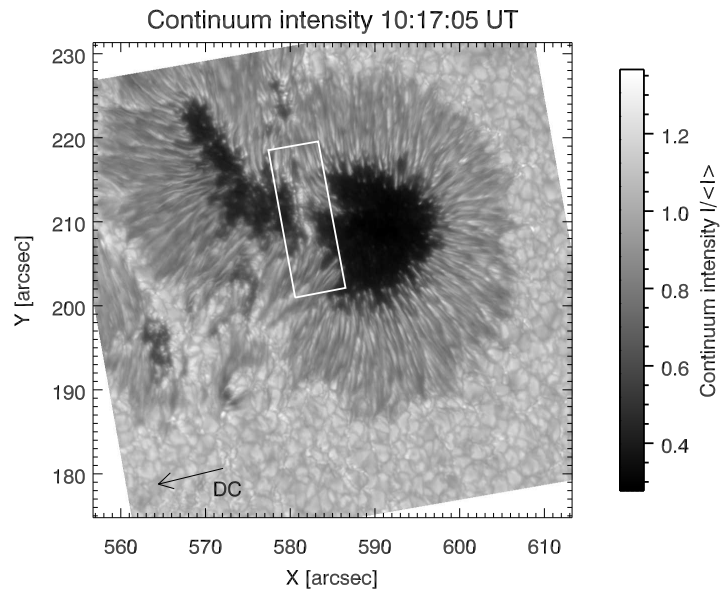}
 \caption{Continuum intensity map of the preceding sunspot of NOAA AR 11263 obtained by CRISP at the Fe I 630.15 nm line on $6$ August 2011 at 10:17 UT. The solid box indicates the LB's FOV analyzed in the text. The arrow points in the direction of disk center.}
 \label{fig3}
\end{figure}

\begin{figure}
   \centering
 \includegraphics[trim=0 0 0 0, clip, scale=0.5]{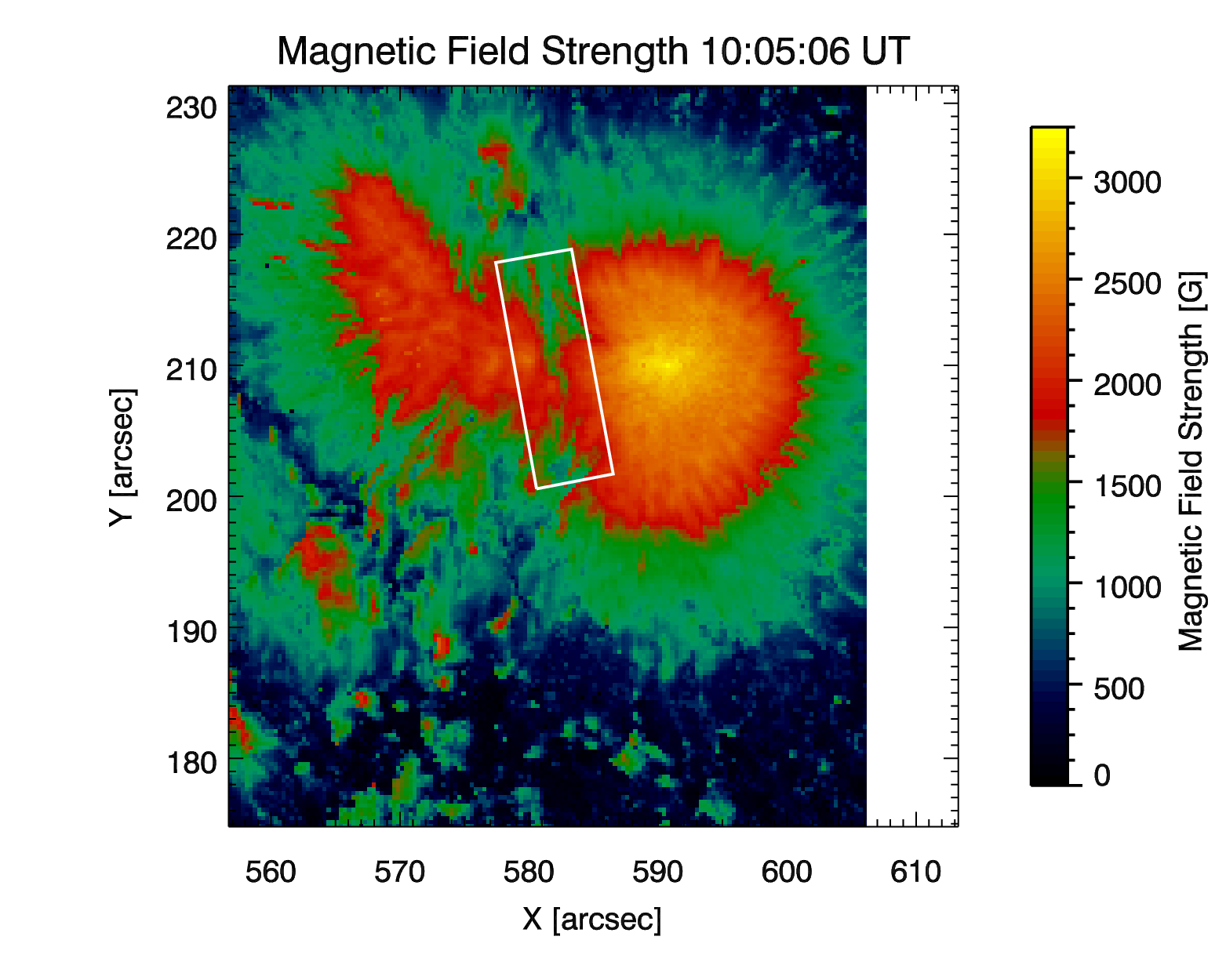}    
 \includegraphics[trim=0 0 0 0, clip, scale=0.5]{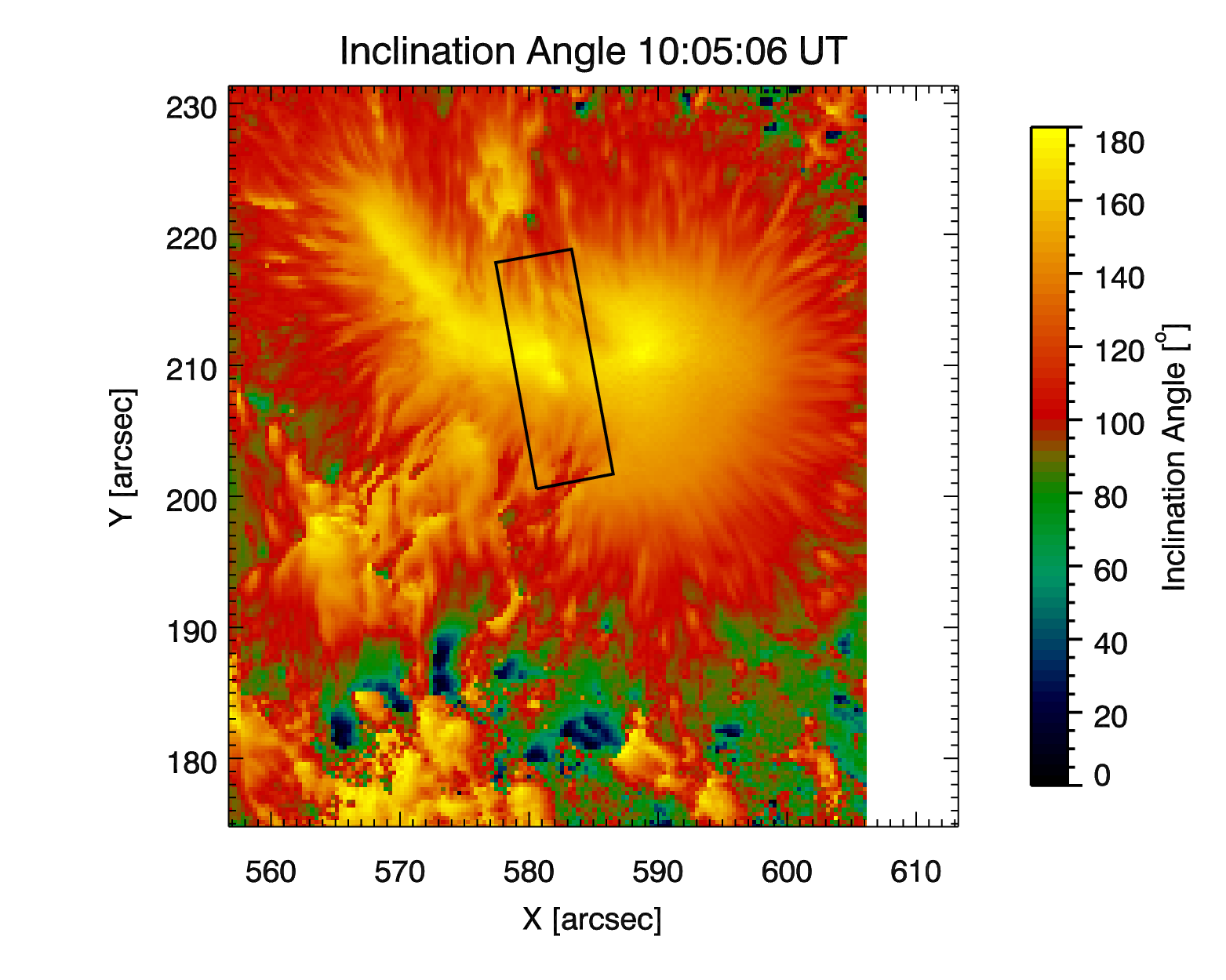}
 \caption{Maps of  the magnetic field strength and inclination angle in the local solar frame coordinate obtained from the inversion of both the Fe I 630.15 nm and 630.25 nm line datasets taken by Hinode on $6$ August 2011 at 10:05:06 UT. The region with $X>606\farcs$ is not covered by Hinode observations. In the bottom panel $0^{\circ}$ and $180^{\circ}$ correspond to the directions of radially outward and inward magnetic field, respectively. In each map the solid box indicates the LB's FOV analyzed in the text.}
 \label{fig4}
\end{figure}

\newpage
\begin{figure}
  \centering
  \includegraphics[trim=0 40 110 450, clip, scale=0.55]{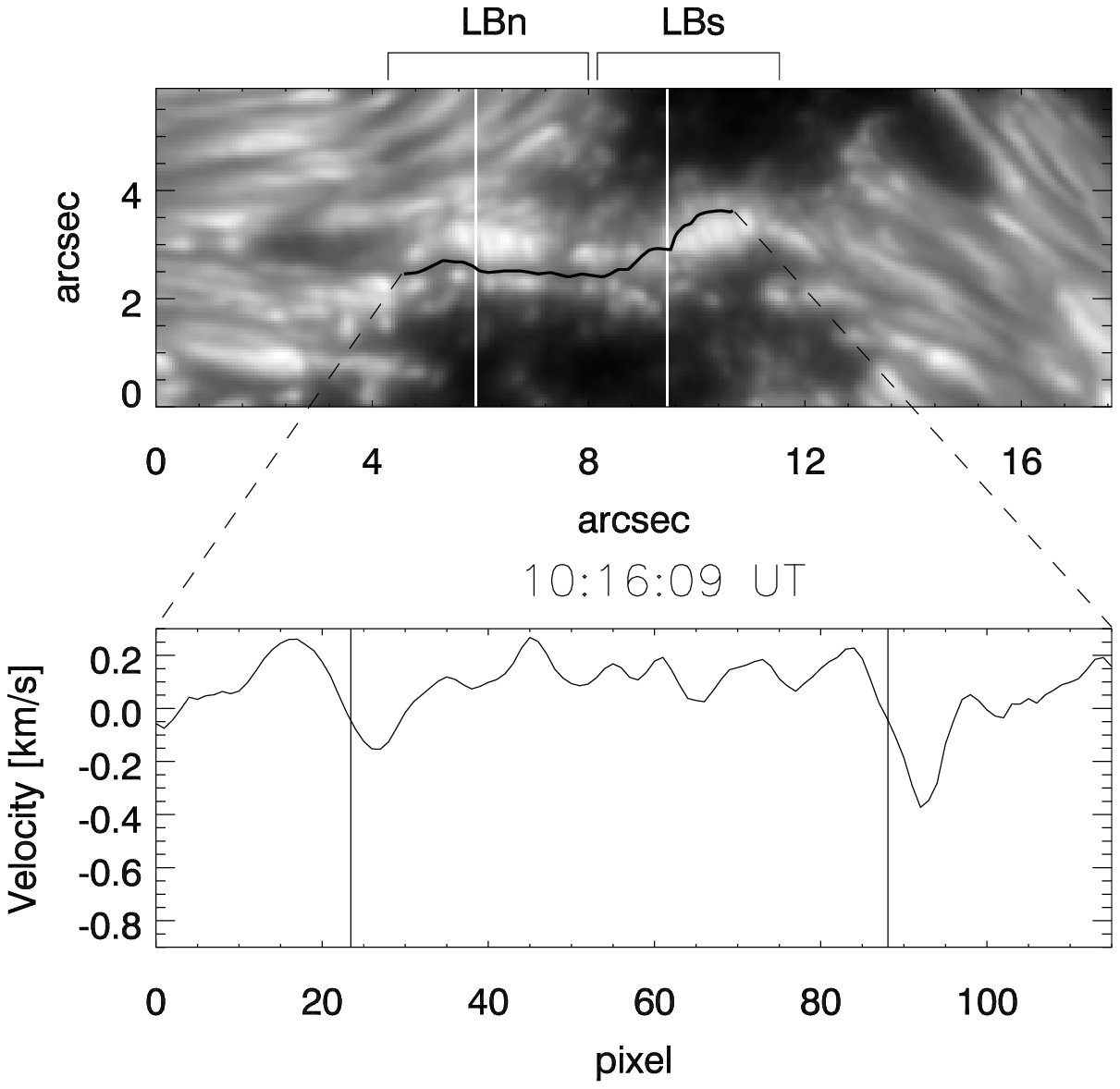}\\
  \includegraphics[trim=0 210 110 450, clip, scale=0.55]{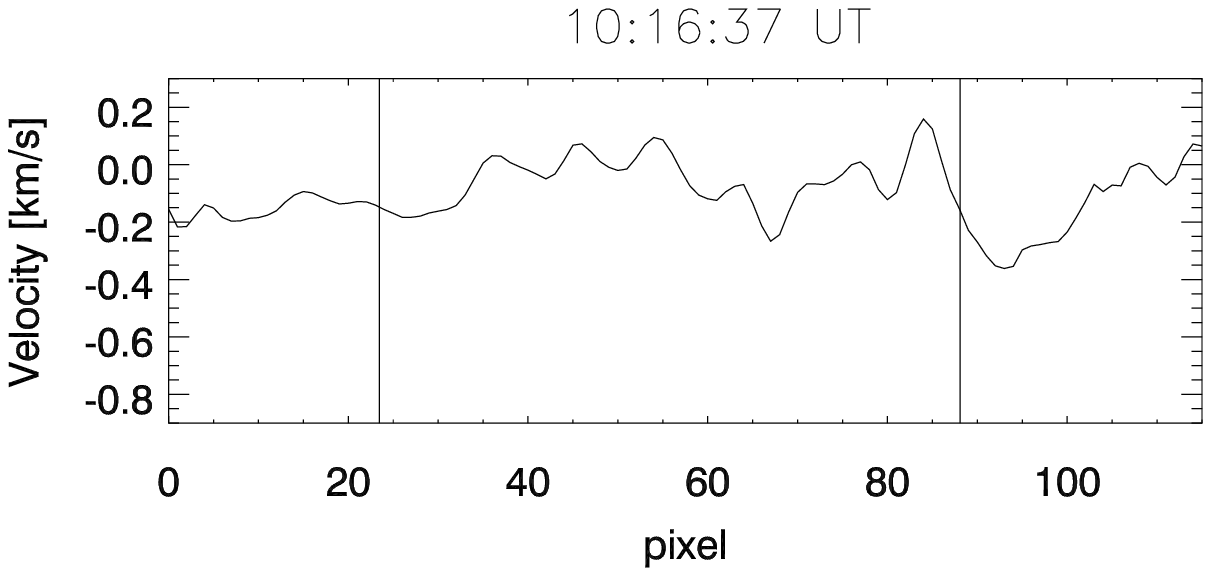}\\
  \includegraphics[trim=0 210 110 450, clip, scale=0.55]{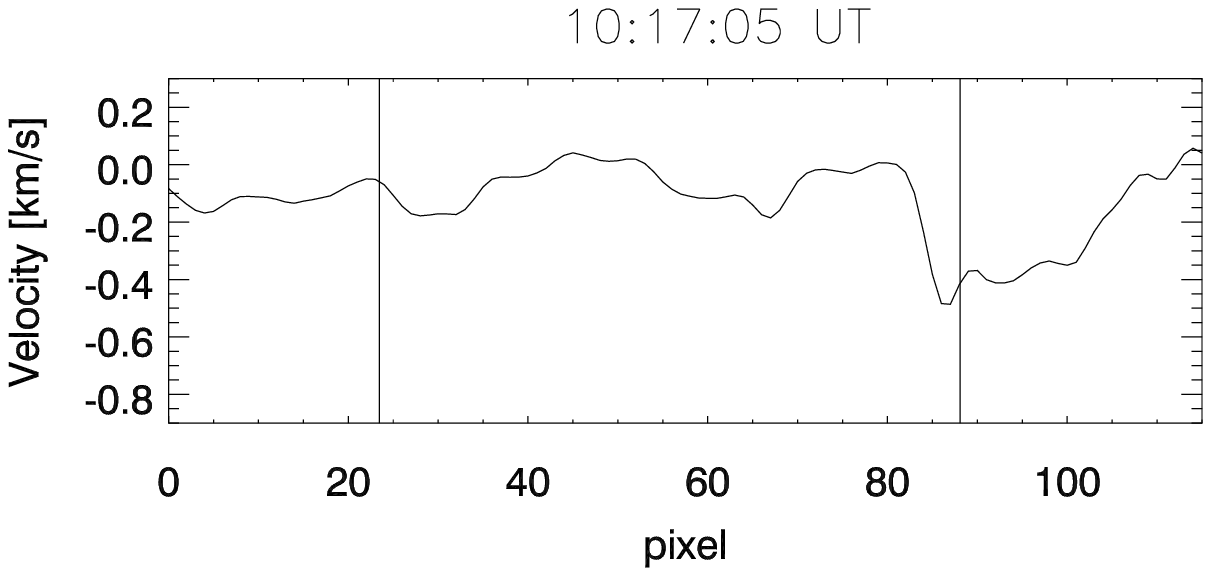}\\
  \includegraphics[trim=0 210 110 450, clip, scale=0.55]{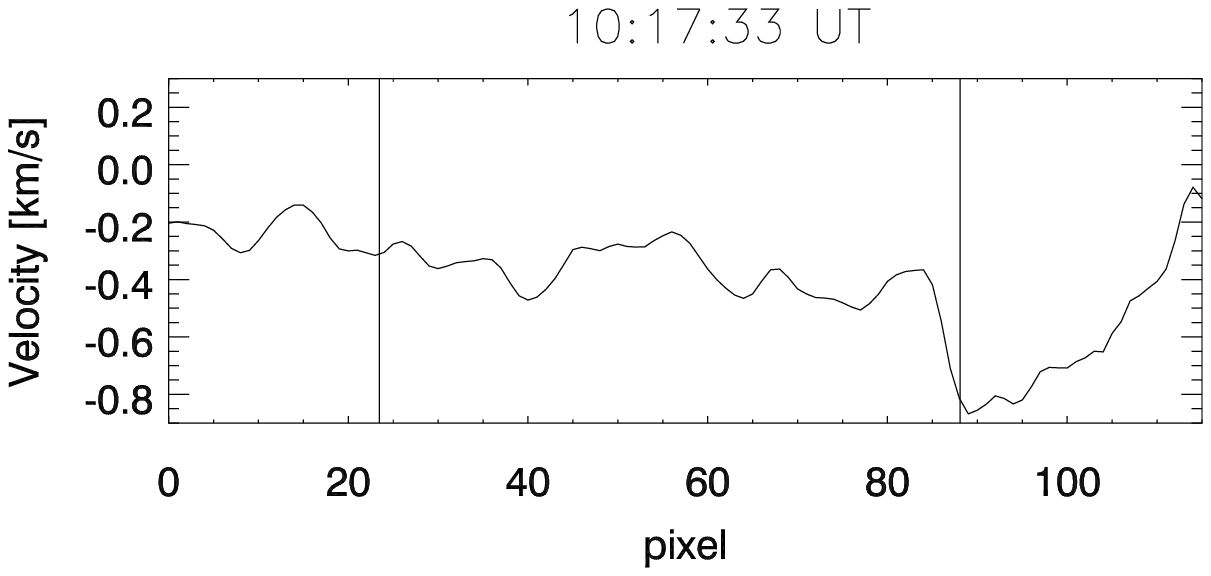}\\
  \includegraphics[trim=0 210 110 450, clip, scale=0.55]{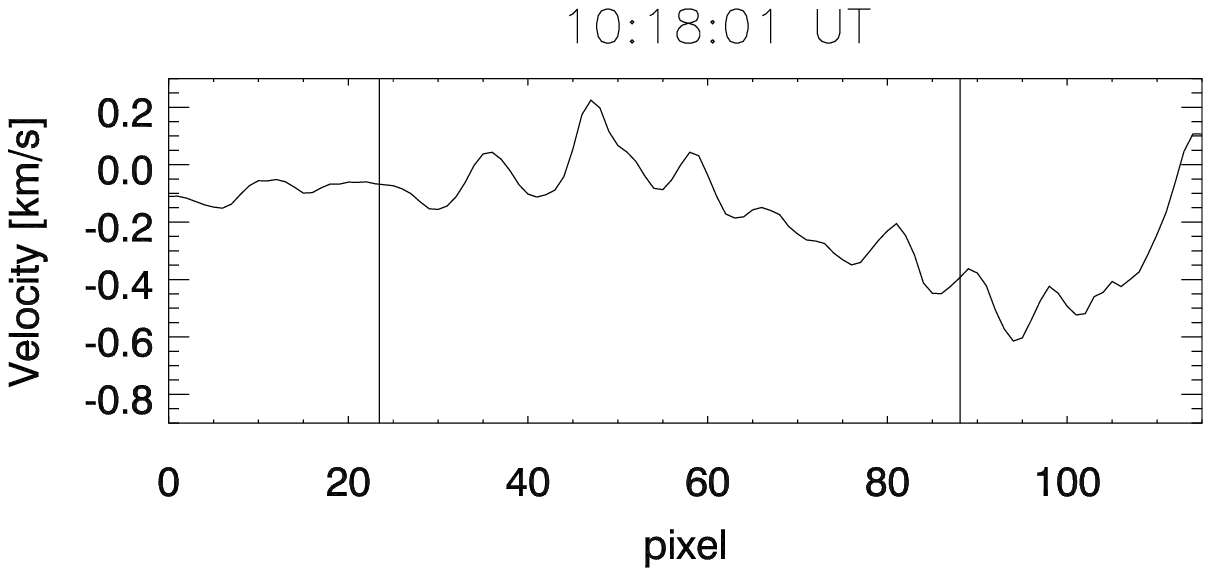}
 \caption{Continuum intensity map from CRISP observations of the LB with an overplotted black line indicating the position of the dark lane studied in the text (top panel) and the LOS plasma velocity values (bottom panels) estimated along the dark lane highlighted on the intensity map. North is on the left of the top image. The two white lines in the top panel and the two black lines in the bottom panel correspond to the lines 1 and 2 reported in Figure \ref{fig6}. Positive (negative) values in velocity indicate motions away from (toward) the observer.}
 \label{fig5}
\end{figure}

\begin{figure}
   \centering
   \includegraphics[scale=0.6]{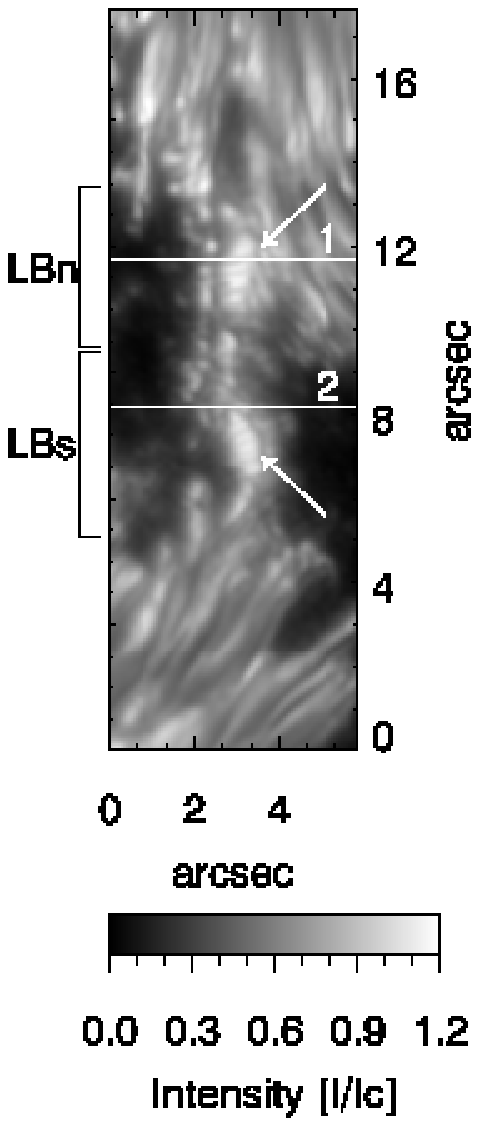}
   \includegraphics[scale=0.6]{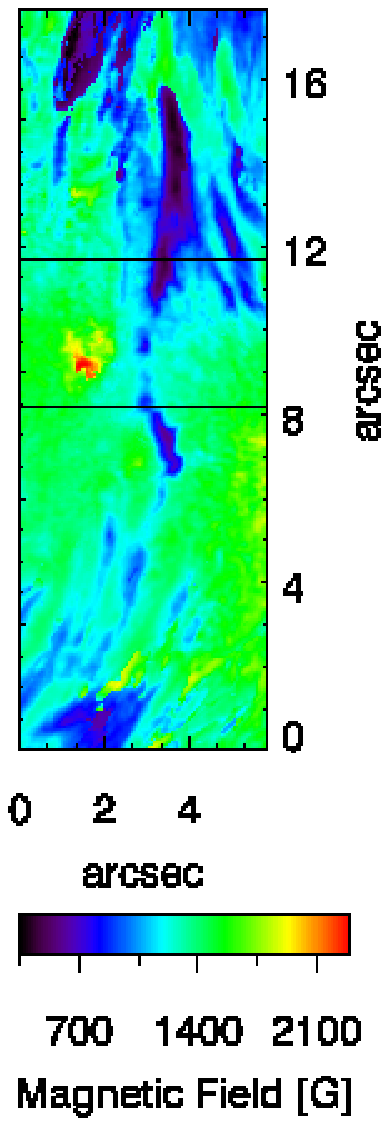} \\
   \includegraphics[trim=0 0 0 50, scale=0.6]{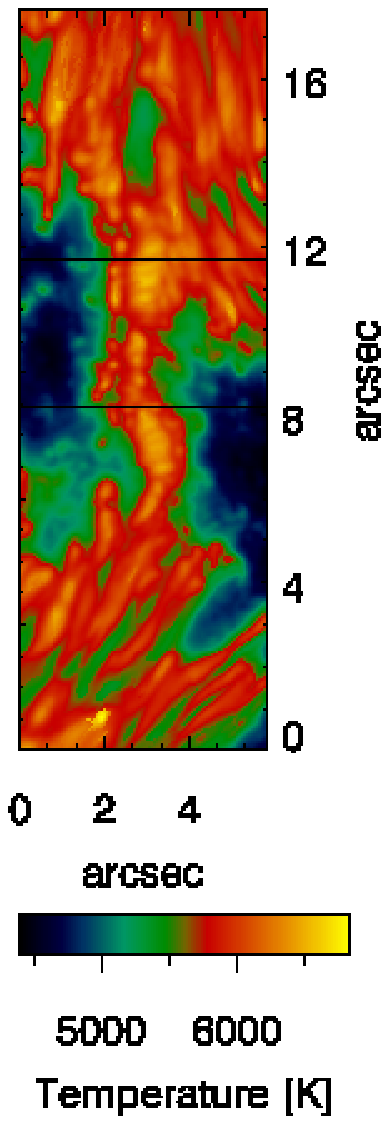}
   \includegraphics[trim=0 0 0 50, scale=0.6]{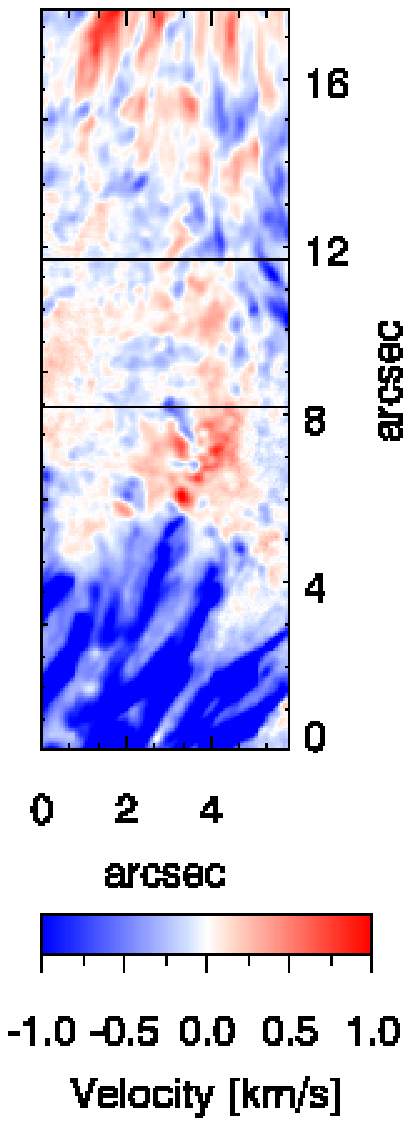}   
 \caption{From top-left, clockwise: maps of the normalized continuum intensity, magnetic field strength, LOS velocity and temperature of the region containing the LB, at 10:17 UT (CRISP dataset). The maps of the magnetic field strength and temperature have been obtained from the SIR inversion of the Stokes profiles along the Fe I 630.15 nm and 630.25 nm lines. In the LOS velocity map, derived from Gaussian fits, positive (negative) values correspond to motion away from the observer (toward the observer).
 The white lines in the intensity map and the dark lines in the other maps correspond to the pixels studied in Figure \ref{fig7}. In each map, line 1 indicates a region of LB$_{n}$ and line 2 indicates a region of LB$_{s}$.}
\label{fig6}
\end{figure}

\begin{figure}
   \centering
   \includegraphics[scale=0.5]{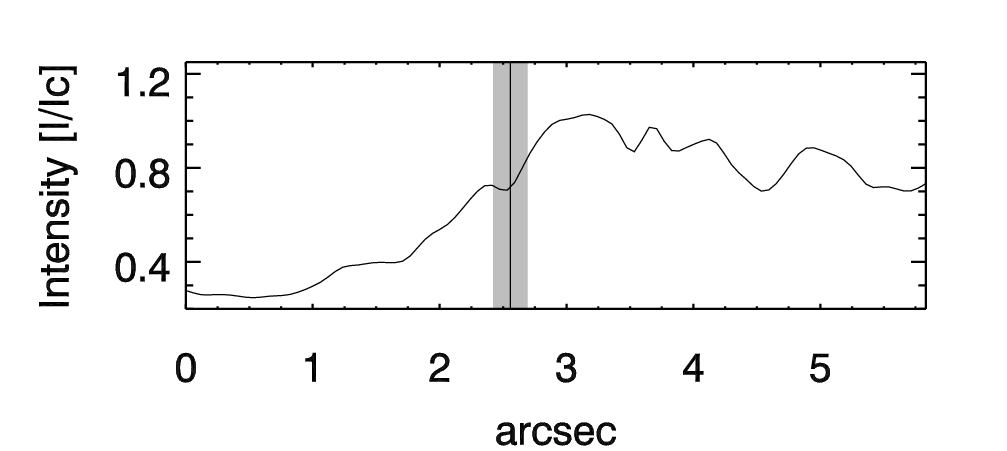}
   \includegraphics[scale=0.5]{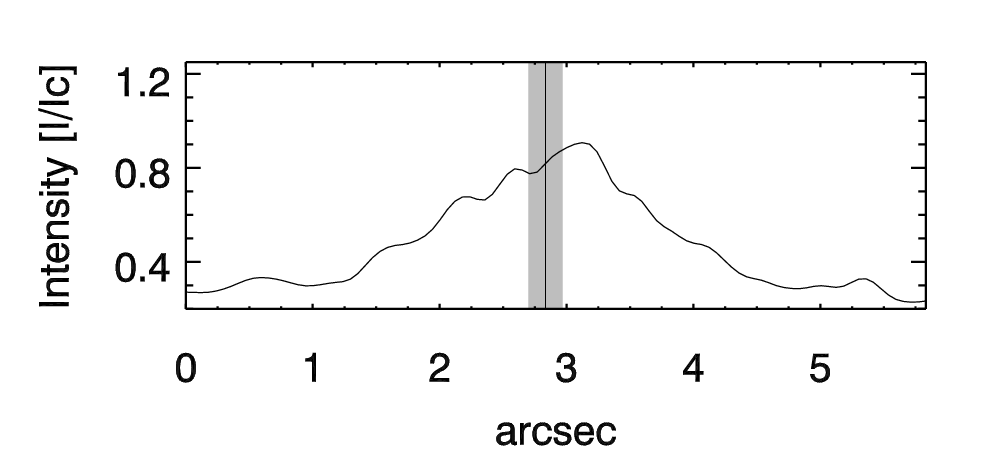}\\
   \includegraphics[scale=0.5]{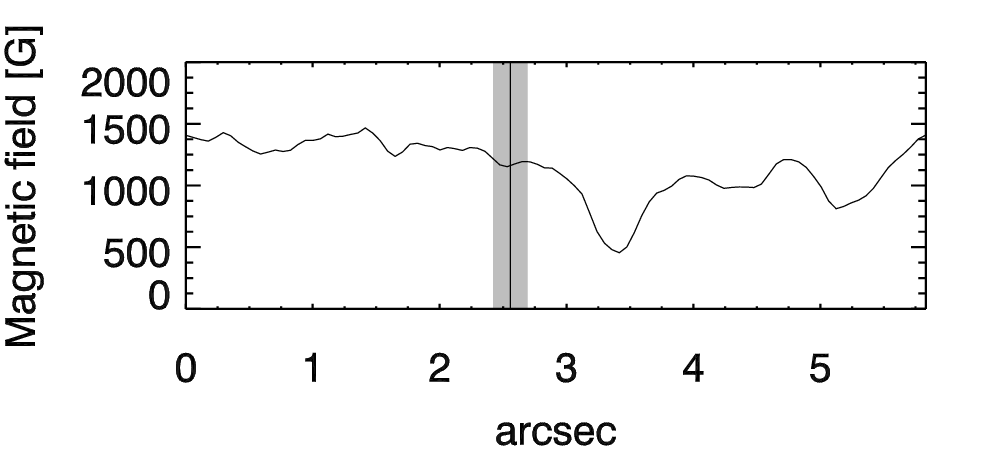}
   \includegraphics[scale=0.5]{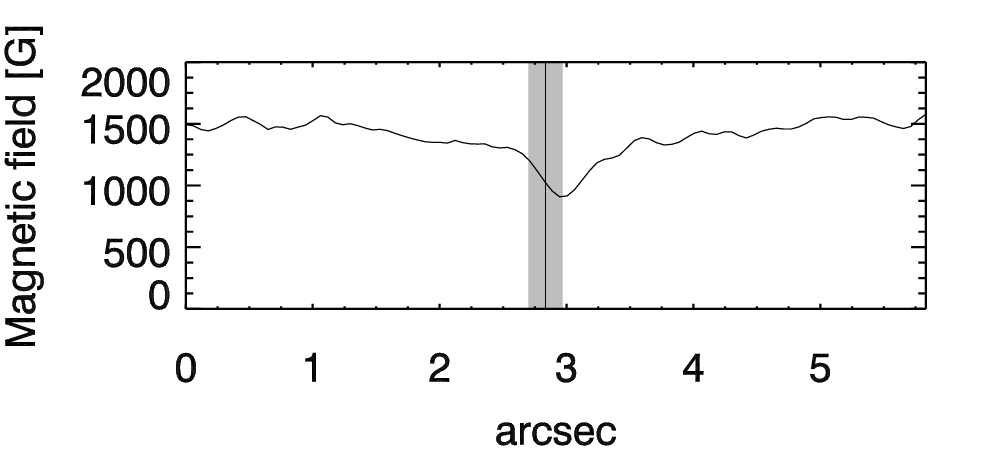}\\
   \includegraphics[scale=0.5]{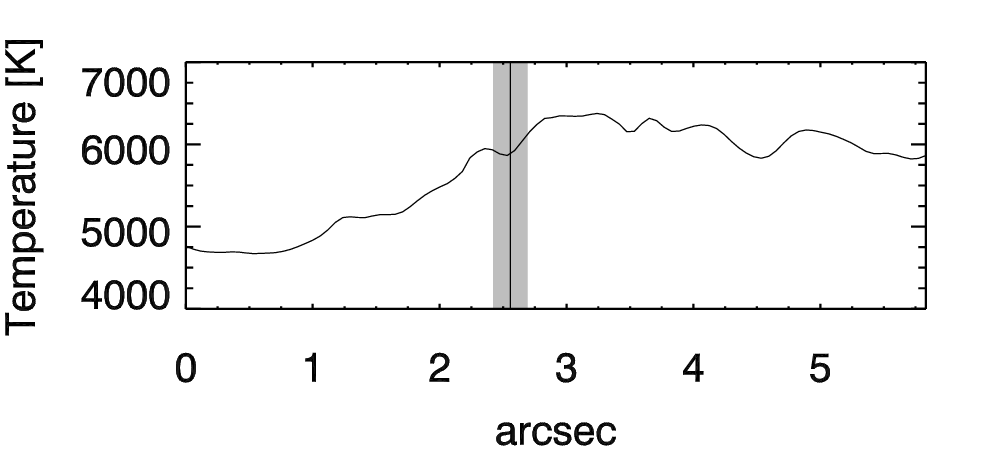}
   \includegraphics[scale=0.5]{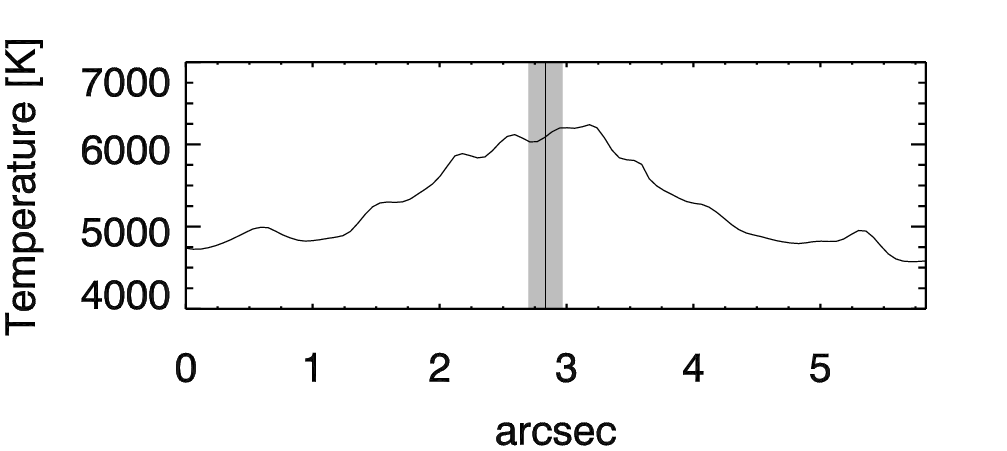}\\
   \includegraphics[trim=0 -60 0 0,scale=0.5]{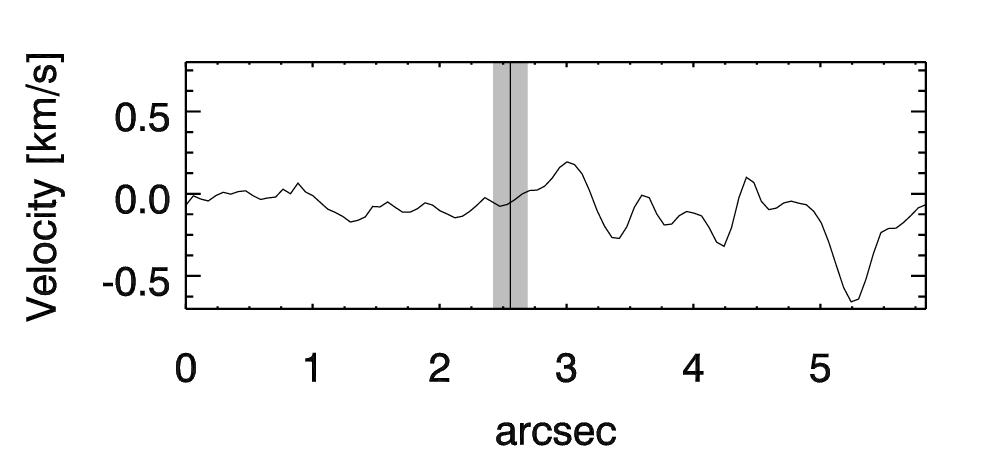}
   \includegraphics[trim=0 -60 0 0,scale=0.5]{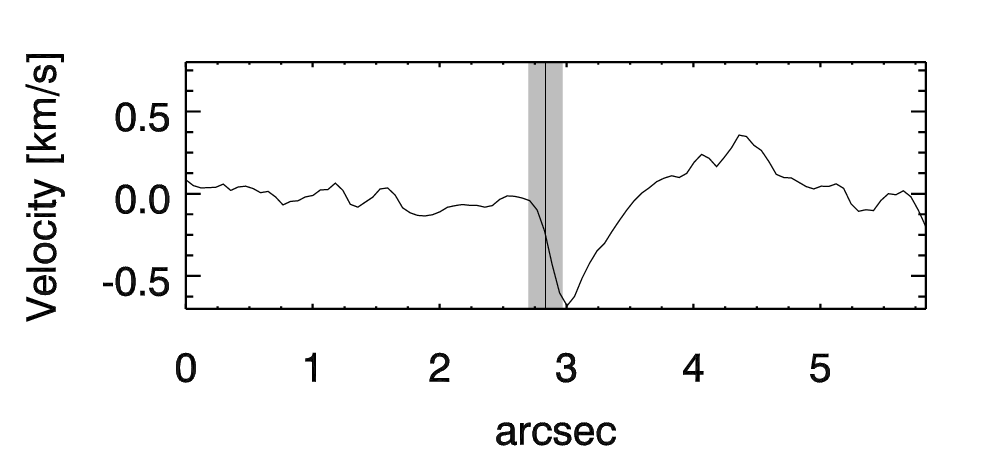}\\
      \caption{Left column: from the top to the bottom, plots of intensity, magnetic field strength, temperature, and LOS velocity along line 1 (see Figure \ref{fig6}). Right column: the same parameters along line 2 (see Figure \ref{fig6}). In these plots the black vertical line shows the dark lane position and the grey line shows its width.} 
    \label{fig7}
      \end{figure}

\begin{figure}
   \centering
   \includegraphics[scale=0.6]{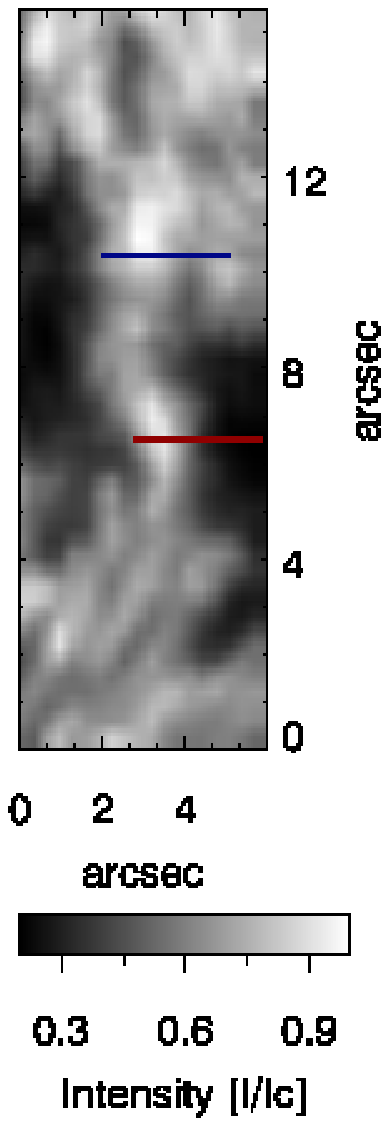}
   \includegraphics[scale=0.6]{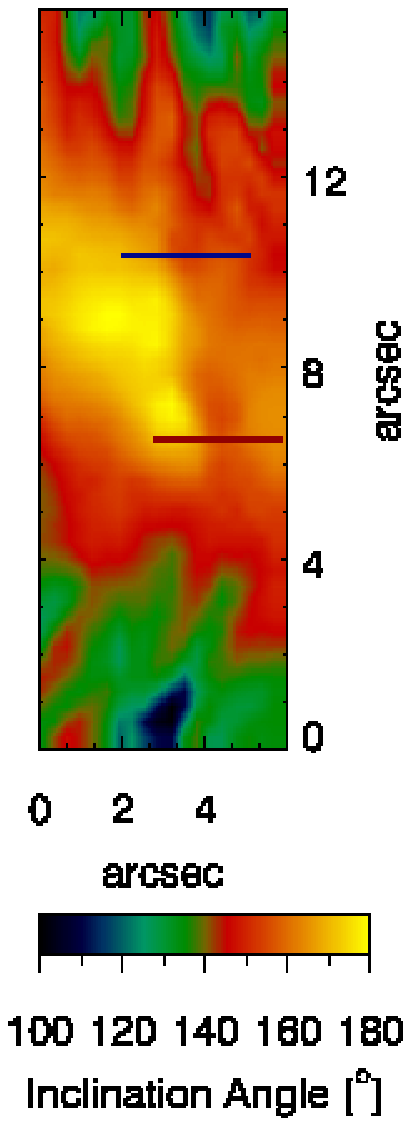}
 \caption{Maps of the intensity (left panel) and inclination angle (right panel) of the region containing the LB (see the solid box in Figure \ref{fig3} indicating the analyzed FOV for the LB deduced from Hinode/SP observations begun at 10:05 UT). The blue and red segments correspond to a part of the white lines 1 and 2, respectively, shown in Figure \ref{fig6} (top-left panel).}
\label{fig8}
\end{figure}

\newpage
\begin{figure}[h]
   \centering
  \includegraphics[scale=0.4]{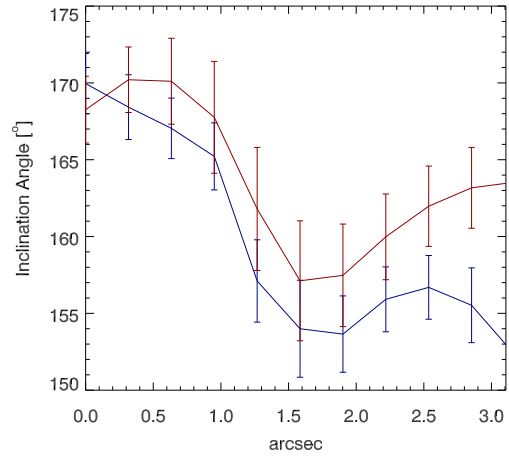}
  \caption{Plot of the inclination angle along the blue and red lines of Figure \ref{fig8} and the respective error bars, corresponding to the $1\sigma$ uncertainty. The blue plot corresponds to the inclination angle along the LB$_{n}$ located between an umbral and a penumbral zone. The red plot corresponds to the inclination angle along the LB$_{s}$ located between two umbral cores.}
\label{fig9}
\end{figure}

\begin{figure}
   \centering
   \includegraphics[trim=0 0 0 0, scale=0.9]{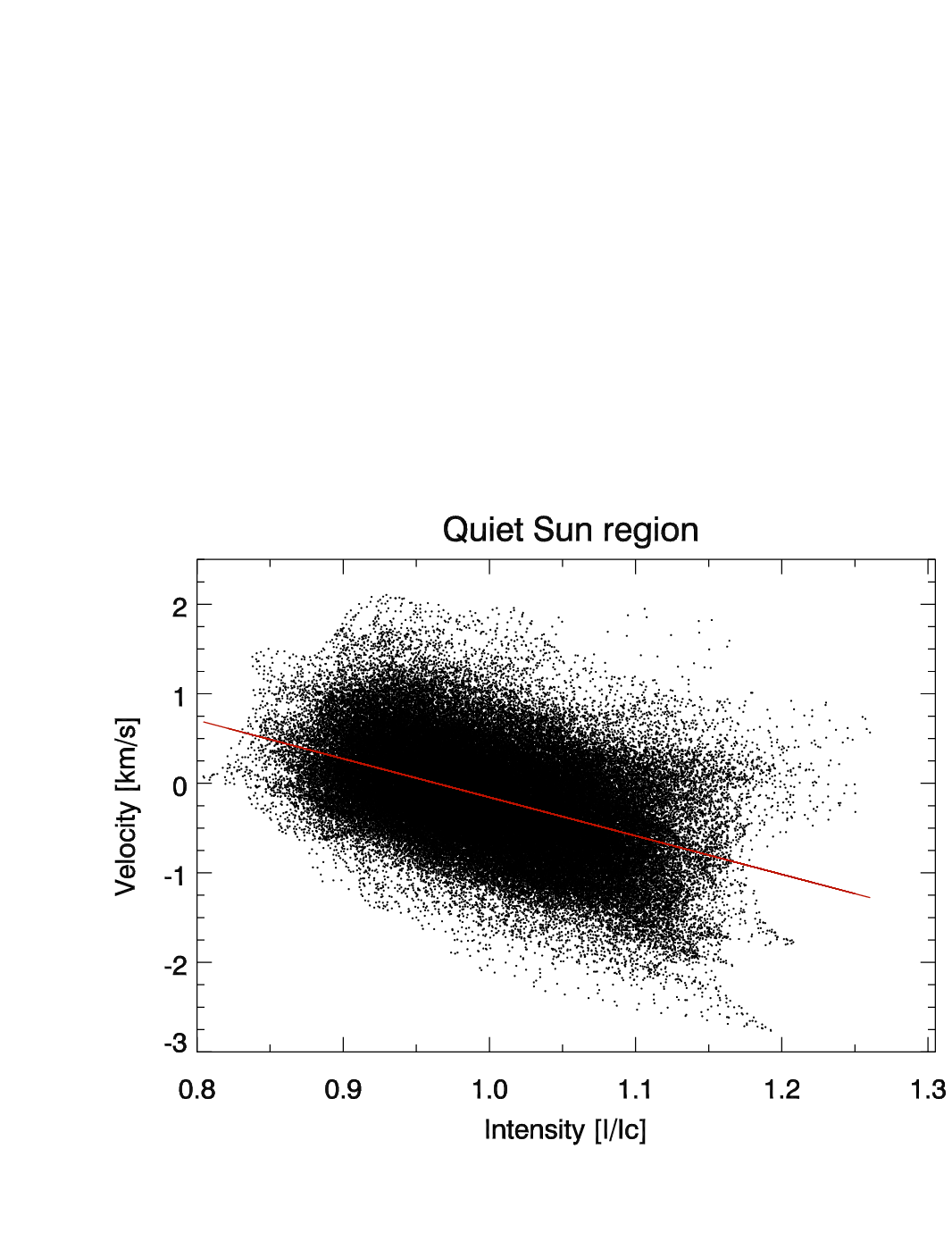}\\
   \includegraphics[trim=0 0 0 150, scale=0.9]{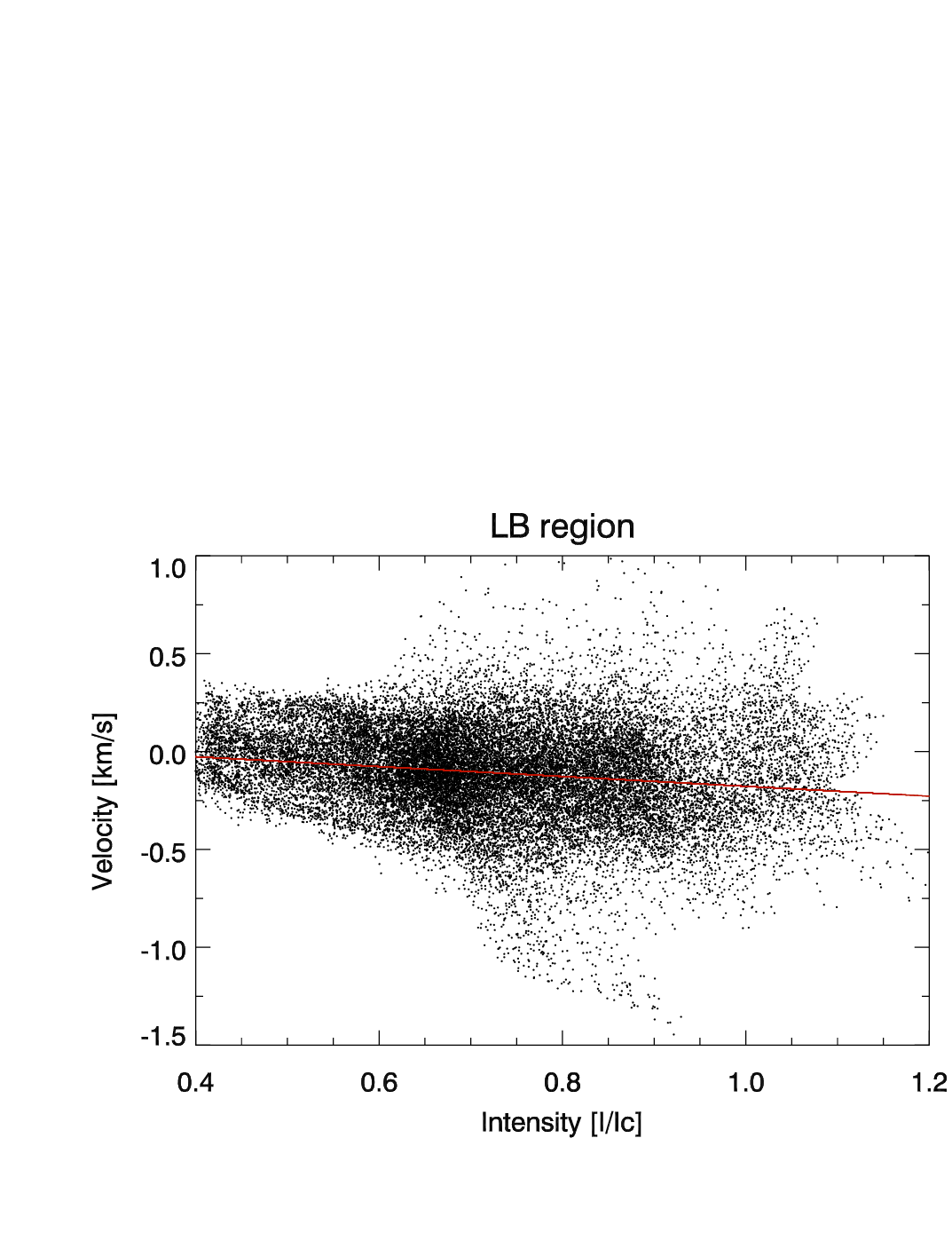}
  \caption{Scatter plots of intensity and LOS velocity in a region of the quiet Sun (top panel) and in the region of the LB (bottom panel). The data plotted in the latter plot include values of the five sequences studied in Figure \ref{fig5}. The red line in each plot represents the linear fit of the dataset analysed.}
\label{fig10}
\end{figure}

\end{article} 
\end{document}